# AnisoCADO: a python package for analytically generating adaptive optics point spread functions for the Extremely Large Telescope

**Kieran Leschinski**[1] **and Eric Gendron**[2]

**1** Department of Astrophysics, University of Vienna  **2** LESIA, Observatoire de Paris, Université PSL, CNRS, Sorbonne Université, Université de Paris

## Summary

AnisoCADO is a Python package for generating images of the point spread function (PSF) for the European Extremely Large Telescope (ELT) (Gilmozzi & Spyromilio, 2007). The code allows the user to set many of the most important atmospheric and observational parameters that influence the shape and strehl ratio of the resulting PSF, including but not limited to: the atmospheric turbulence profile, the guide star position for a single conjugate adaptive optics (SCAO) solution, differential telescope pupil transmission, etc. Documentation can be found at https://anisocado.readthedocs.io/en/latest/

## Statement of need

### Adaptive optics are mandatory for the next generation of ground-based telescopes

The larger the telescope aperture, the greater the spatial resolution of the observations. For space-based telescopes this statement is always true. However the resolution of ground based telescopes is limited by the blur caused by turbulence in the atmosphere - known as "atmospheric seeing". This blurring can be (mostly) removed by measuring the deformation of the wavefront of the incoming light, and applying an equal and opposite deformation to the surface of one or more of the mirrors along a telescope's optical path. The current fleet of large (8-10m) telescopes were built to primarily operate at the edge of the natural seeing limit (FWHM~0.5 arcseconds at $\lambda=1\mu$m). Over the last two decades some have received upgrades in the form of active and adaptive mirrors, in order to achieve up to a 20x increase in resolution afforded by the physical diffraction limit of a 10m primary mirror (FWHM~0.03 arcseconds at $\lambda=1\mu$m). The next generation of "extremely large" telescopes will have primary mirrors on the order of 30-40m, with theoretical diffraction limits on the order of 100x smaller than the natural seeing limit in the near infrared range. In order for these telescopes to resolve structures at scales similar to the diffraction limit, they must, by design, include adaptive optics systems.

### Diffraction limited point-spread-functions are complex beasts

The PSF of an optical system is the description of the spatial distribution of light from an infinitely small point source after passing through an optical system (e.g. layers of the atmosphere, mirrors of a telescope). Due to the random nature of atmospheric turbulence, the averaged long-exposure PSF of a star in a seeing-limited observation is well approximated by a smooth, yet very wide, Gaussian-like function. The PSF of a diffraction limited telescope system using an adaptive-optics correction is a complex, sharp function that depends on a veritable zoo of atmospheric, observational, and technical parameters. From an astronomer's point of view, the consequences of a poor adaptive optics solution means the difference between a successful and a failed observation run. Therefore it is imperative that the consequences of such large variations in the PSF are accounted for in advance, in order to guarantee the feasibility of the proposed programs.

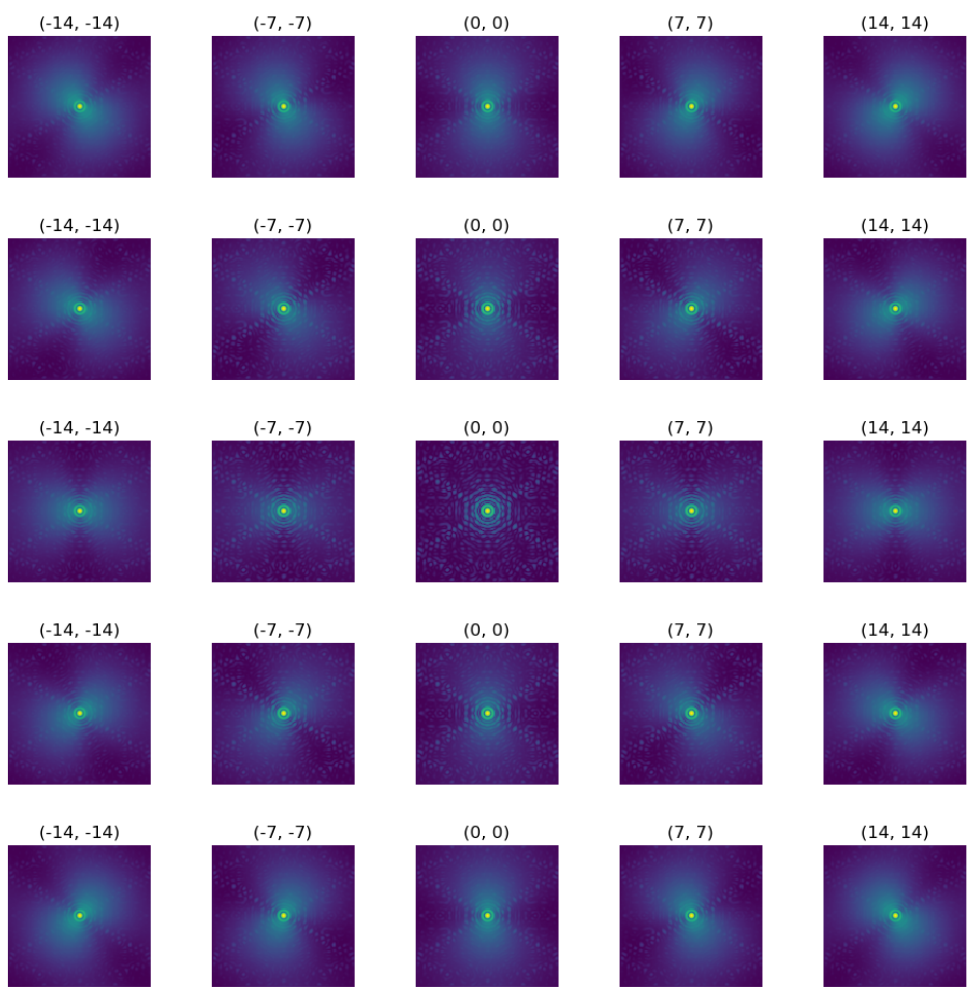

**Figure 1:** A grid of Ks-band (~2.15$\mu$m) PSFs for a range of distances from the natural guide star. The PSFs were generated using the ESO median turbulence profile.

## AnisoCADO - Anisoplanatism for MICADO

AnisoCADO (Anisoplanatism for MICADO) is a package for generating images of the point spread function for a given set of atmospheric, observational, and technical conditions. It achieves this in Fourier space, by combining a series of optical transfer functions (OTF), each of them corresponding to an effect that contributes to the residual phase error (e.g. deformable mirror fitting error, system temporal error, system aliasing). The product of the individual OTFs is applied to the OTF of the telescope optical system, which includes all the typical features of the telescope pupil (spiders, central obscuration, etc). Lastly a fourier transfor-mation is applied to the full system OTF. The resulting image is the expected PSF for a long exposure (>10s) on-axis observation at the given wavelength. For single-conjugate adaptive optics (SCAO) modes, the field PSF degrades as distance from the guide star increases. This effect is taken into account by introducing a supplementary OTF specific to anisoplanatism, which varies with the off-axis distance. Figure 1 shows how the PSF changes with distance from an on-axis guide star.

For a more detailed discussion of the mathematics behind anisoplanatism in the context of the ELT, the reader is referred to Clénet, Gendron, Gratadour, Rousset, & Vidal (2015).

**Inputs**

The final ELT PSF is the combination of many factors. The vast majority of these are irrelevant for the casual user. AnisoCADO therefore provides three preset options, corresponding to the standard ESO best (first quartile), median, and worst (fourth quartile) turbulence conditions (Farley et al., 2018). All other parameters are initialised with default values. For the case of a MICADO SCAO system (Clénet et al., 2014; Davies et al., 2018) PSFs can be generated for multiple guide star offsets without needing to re-make the OTFs by using the special class method `.shift_off_axis(dx, dy)`.

For more detailed use cases, the following parameters are available to the user:

| Atmosphere | Observation | Telescope and AO system |
|---|---|---|
| Turbulence profile | Natural guide star position | Pupil image |
| Altitude of turbulent layers | Central wavelength | 2D pupil transmissivity |
| Strength of turbulent layers | Pupil rotation angle | Empty mirror segments |
| Wind speed | Zenith distance of observation | Plate scale |
| Seeing FWHM at 500nm | | Residual wavefront errors |
| Fried parameter | | AO sampling frequency |
| Outer scale | | AO loop delay |
| | | Interactuator distance |

**Outputs**

AnisoCADO is easily integrated into the standard astronomer's toolbox. PSF images generated by AnisoCADO can be output as either `numpy` arrays (van der Walt, Colbert, & Varoquaux, 2011), or standard `astropy.io.fits.ImageHDU` objects (Astropy Collaboration et al., 2018). The latter can be written to file using the standard `astropy` syntax.

As AnisoCADO was written to support the development of the MICADO instrument simulator (Leschinski et al., 2016, 2019), it is possible to generate `FieldVaryingPSF` objects using the helper functions in the `misc` module. Such files are also compatible with the generic instrument data simulator framework, ScopeSim.

**Basic Example**

The AnisoCADO API is described in the online documentation, which can be found at: https://anisocado.readthedocs.io/. For the purpose of illustration, the following 5 lines were used to generate the grid of PSFs in Figure 1.

```
import numpy as np
from anisocado import AnalyticalScaoPsf
```

```
psf = AnalyticalScaoPsf()
psf_grid = []
for x, y in np.mgrid[-14:15:7, -14:15:7].flatten().reshape((2, 25)).T:
    psf.shift_off_axis(x, y)
    psf_grid += [psf.kernel]
```

# Acknowledgments

AnisoCADO depends on the following packages: Numpy (van der Walt et al., 2011), Matplotlib (van der Walt et al., 2011), Astropy (Astropy Collaboration et al., 2018).

This project was funded by project IS538004 of the Hochschulraum-strukturmittel (HRSM) provided by the Austrian Government and administered by the University of Vienna.